\documentclass[12pt]{article}
 \usepackage{graphicx} 
 \usepackage{epsfig}
 \usepackage{bm}
\def\email#1{\date{\tt#1}}
\def\address#1{\par\noindent#1\smallskip}

\begin{document}

\title{Super-elastic collisions of thermal activated nanoclusters}
\author{Hisao ~Hayakawa$^{\dagger}$, \and Hiroto ~Kuninaka$^{\S}$ }

\email{$\dagger$hisao@yukawa.kyoto-u.ac.jp,$\S$kuninaka@phys.chuo-u.ac.jp}

\maketitle
\thispagestyle{empty}

\begin{abstract} 
Impact processes of nanoclusters subject to thermal fluctuations are 
investigated, theoretically. 
In the former half of the paper, we discuss the basis of 
quasi-static theory.  In the latter part, 
we carry out the molecular dynamics simulation of collisions between 
two identical nanoclusters, and report some statistical properties of impacts of nanoclusters. 
\end{abstract}

\section{Introduction}
The initial kinetic energy 
of colliding bodies is distributed into the internal degrees of freedom in an inelastic collision. 
Such a collision is characterized by the restitution coefficient $e \equiv  V^{'}/V$, 
where $V$ and $V'$ are, respectively, the relative colliding speed and the relative rebound speed.
Although it is believed that the restitution coefficient $e$ satisfies $e<1$ for impacts of macroscopic bodies,
 the anomalous impact with $e>1$ is possible 
in some special situations for small bodies.
Indeed, the prohibition of $e>1$ is
 originated from the second law of thermodynamics\cite{maes-tasaki,tasaki-jsp}, but some terms which disappear in
the  thermodynamic limit play important roles in the description of small systems. 
It should be noted that 
the restitution coefficient projected 
into the normal direction of the collision 
can easily exceed unity in the case of oblique collisions.\cite{louge,kuninaka_prl}  


The low-speed collisions for macroscopic bodies are believed to be described by 
the quasi-static theory, which is consistent with  
some experimental results.\cite{kuwabara, brilliantov96,morgado} 
However, it is not obvious whether the quasi-static theory is applicable to the impact of nanoclusters. 
Indeed, we expect that the effects of cohesive force among atoms cannot be ignored for such small systems.
Awasthi {\it et al.}\cite{awasthi} 
reported that the dependence of the restitution coefficient $e$ on the impact speed 
for nanoclusters, which contain adhesions,
differs from the prediction from the quasi-static theory based on their molecular dynamics simulation (MDS).
In a recent paper, Brilliantov {\it et al.} extend the quasi-static theory to the theory of cohesive 
collisions.\cite{brilliantov-adh} 

The physics of nanoclusters is one of hot subjects. 
The MDS  is a standard tool 
to investigate collisions of nanoclusters such as fulleren. 
Some of such studies  focus on fragmentations and coalescences
after binary collisions of nanoclusters\cite{kalweit,lewis,knospe}. 
The other studies discuss  the collisions of a cluster with a substrate\cite{awasthi},  
the erosion process on a diamond surface\cite{yamaguchi}, 
and the fragmentation pattern of clusters\cite{tomsic}. 
So far, we do not know any paper to investigate the effects of 
thermal fluctuations on collisions of nanoclusters except for our preliminary report.\cite{kuninaka-preprint}

In this paper, we perform the MDS of 
colliding clusters to investigate the effect of 
thermal fluctuations.  This is the extension of our previous work.\cite{kuninaka-preprint}
In the first part,
 we review what model is adequate to describe the collision of nanoclusters. 
In the next section, we introduce the generalized Langevin equation and the fluctuation-dissipation relation in the first kind.
In section 3, we  calculate the velocity autocorrelation function (VACF)  of a lattice model to apply it to a system of a nanocluster.
Then, we justify the quasi-static theory to describe the collisions of nanoclusters.
In the second part, we show the detailed results of our numerical simulations. 
In section 4, we introduce our numerical model of the MDS.
In section 5, 
we explain the results of our simulation, which consist of four subsections.
In the first subsection, we check the relaxation of VACF in our model.
In the next subsection, we demonstrate sequential snapshots of  collisions between two identical nanoclusters.
In section 5.3, we compare our numerical result of the impact speed dependence of  the restitution coefficient
with the quasi-static theory of cohesive or noncohesive collisions.
In section 5.4, we show the frequency distribution functions of the restitution coefficient and their dependence on cohesive force between atoms.
We also show the probabilities to appear four categories in  our simulation when the cohesive parameter is finite.
In section 6, we discuss and summarize our results. 

\section{The Langevin equation}
It is well known that we can formally rewrite the Newtonian equation of motion as the generalized Langevin equation for the `slow' variable.
When we consider the motion of colliding a pair of small clusters, it is natural to adopt the relative velocity $\bm{v}$
 between the center of mass of each cluster as the 'slow'  variable\cite{zwanzig}. 
We should note that the center of mass is characterized by the total mass of one cluster, while
each element of the cluster can be characterized by the mass for the element.  Thus, the effective mass of the center of mass is much larger than the mass of the element.
Thus, the generalized Langevin equation is given by
\begin{equation}
\frac{d\bm{v}}{dt}=-\int_{-\infty}^tdt'\gamma(t-t') \bm{v}(t')+\bm{\theta}(t)+\frac{\bm{F}(t)}{M} ,
\end{equation}
where $\gamma(t)$, $\bm{\theta}(t)$, $M$, and $\bm{F}(t)$ are the memory kernel, the fluctuating force from the fast oscillations, the reduced mass of 
two clusters, and the systematic force acting on the centers of mass, respectively. 
The fluctuation force $\bm{\theta}(t)$ is believed to be unimportant for the impact problem
of two clusters. The systematic force $\bm{F}(t)$ may be approximated by the Hertzian contact force. 
From the fluctuation-dissipation relation in the first kind, the Laplace transform $\hat\gamma(\omega)\equiv \int_0^{\infty}dt\gamma(t) e^{-i\omega t}$ satisfies
\begin{equation}\label{fdr}
(i\omega+\hat\gamma(\omega))^{-1}=\frac{M}{T}\int_{-\infty}^{\infty}dt\langle \bm{v}(0)\cdot\bm{v}(t) \rangle e^{-i\omega t},
\end{equation}  
where $T$ is the temperature, and the Boltzmann constant is set to be unity\cite{kubo}.
We should note that $\hat{\gamma}(\omega)$ can be defined as the usual Fourier transform if we assume $\gamma(t)=\gamma(-t)$
for $t<0$.
If the integration of the velocity autocorrelation function (VACF), {\it i.e.} $\int_0^{\infty}dt\langle \bm{v}(0)\cdot\bm{v}(t)\rangle$ is finite, 
$\hat\gamma(\omega\to 0)$ is finite. In this case, the generalized Langevin equation
can be approximated by the Langevin equation with the white noise satisfying $\langle \theta_i(t)\cdot\theta_j(t')\rangle =2T \gamma \delta_{i,j}\delta(t-t')$ where $\theta_i(t)$ is
the $i$-th component of $\bm{\theta}(t)$.
It is obvious that this Langevin equation is  an irreversible equation.  Thus, the behavior of VACF is the most important to characterize the macroscopic dissipation.

\section{The relaxation of the correlation function}

As discussed in the previous section, 
the relaxation of VACF plays a key role for the equilibration process of a system. 
Let us consider the relaxation of VACF in a nanocluster which consists of a regular lattice with equal-mass
atoms.
When the atoms are confined in attractive potential, the excitation from the ground state is characterized by
the harmonic oscillation in a simple cubic lattice.
In this section, we demonstrate that VACF of a uniform system in the simple cubic lattice exhibits the slow 
relaxation proportional to $1/\sqrt{t}$.
The analysis presented here is the straightforward extension of 
the one-dimensional cases.\cite{zwanzig}

Let us consider an infinitely large simple cubic lattice system in which the mass points with mass $m$ connecting with the linear spring whose spring constant is $k$.
The position of each lattice point can be specified by a set of integer $\bm{n}=(n_x,n_y,n_z)$ in this system. 
Introducing the characteristic angular frequency $\omega_0\equiv \sqrt{k/m}$, the equation of motion of the deviation from the equilibrium position 
$\bm{r}_{\bm{n}}(t)$ obeys
\begin{equation}
\ddot{\bm{r}}_{\bm{n}}(t)=-\omega_0^2(6\bm{r}_{\bm{n}}(t)-\sum_{i=1}^6\bm{r}_{\bm{n}+\hat{\bm{e}}_i}(t)) ,
\end{equation}
where $\hat{\bm{e}}_1=(1,0,0)$, $\hat{\bm{e}}_2=(-1,0,0)$, 
$\hat{\bm{e}}_3=(0,1,0)$, $\hat{\bm{e}}_4=(0,-1,0)$, $\hat{\bm{e}}_5=(0,0,1)$ and $\hat{\bm{e}}_6=(0,0,-1)$.

Let us introduce the lattice Fourier transform and the inverse Fourier transform as
\begin{equation}
\hat{\bm{r}}_{\bm{k}}(t)\equiv \sum_{\bm{n}}e^{-i\bm{k}\cdot\bm{n}}\bm{r}_{\bm{n}}(t), \quad
\bm{r}_{\bm{n}}(t)=\frac{1}{(2\pi)^3}\int d\bm{k} e^{i\bm{k}\cdot\bm{n}}\hat{\bm{r}}_{\bm{k}}(t),
\end{equation}
where $\sum_{\bm{n}}=\sum_{n_x=-\infty}^{\infty}\sum_{n_y=-\infty}^{\infty}\sum_{n_z=-\infty}^{\infty}$ and
$\int d\bm{k}=\int_{-\pi}^{\pi}dk_x\int_{-\pi}^{\pi}dk_y\int_{-\pi}^{\pi}$.
Thus, the equation of motion in the Fourier space is given by
\begin{equation}
\ddot{\hat{\bm{r}}}_{\bm{n}}(t)=-4\omega_0^2\sum_{i=1}\sin^2\left(\frac{k_i}{2}\right) \hat{\bm{r}}(t) .
\end{equation}
Furthermore, introducing the Laplace transform $\tilde{\bm{r}}_{\bm{k}}(z)\equiv \int_0^{\infty}dte^{-zt} \bm{\hat{r}}_{\bm{k}}(t)$, we obtain
\begin{equation}
\tilde{\bm{r}}_{\bm{k}}(z)=\frac{ z \hat{\bm{r}}_{\bm{k}}(0)+\dot{\hat{\bm{r}}}_{\bm{k}}(0) }{ z^2+4\omega_0^2\sum_{i=1}^3\sin^2\left(\frac{k_i}{2}\right) }.
\end{equation}

Let us consider the motion of the mass point at the center of the mass in the system. From the relation 
$\bm{r}_{\bm{0}}(t)=\int \frac{d\bm{k}}{(2\pi)^3}\hat{\bm{r}}_{\bm{k}}(t)$, we reach 
\begin{equation}
\bar{\bm{r}}_{\bm{0}}(z)=\frac{1}{(2\pi)^3}\int d\bm{k}\frac{ z \hat{\bm{r}}_{\bm{k}}(0)+\dot{\hat{\bm{r}}}_{\bm{k}}(0) }{ z^2+4\omega_0^2\sum_{i=1}^3\sin^2\left(\frac{k_i}{2}\right) }.
\end{equation}
where we have used $\bar{\bm{r}}_{\bm{n}}(z)\equiv \int_0^{\infty}dte^{-zt} \bm{r}_{\bm{n}}(t)$.

Here, VACF at the center of the mass is defined by
\begin{equation}\label{vacf-com}
\phi(t)\equiv \langle \dot{\bm{r}}_{\bm{0}}(0)\cdot\dot{\bm{r}}_{\bm{0}}(t)\rangle ,
\end{equation}
where $\langle \cdots \rangle$ represents the ensemble average over the different initial conditions. We assume that the initial condition satisfies
\begin{equation}
\langle \dot{\bm{r}}_{\bm{m}}(0)\cdot\dot{\bm{r}}_{\bm{n}}(0)\rangle =
\phi(0)\delta_{\bm{m},\bm{n}}=
\frac{3T}{m} \delta_{\bm{m},\bm{n}}
, \quad \langle \bm{r}_{\bm{n}}(0)\cdot\dot{\bm{r}}_{\bm{0}}(0)\rangle =0.
\end{equation}
Then, the Laplace transform 
$\tilde{\phi}(z)\equiv \int_0^{\infty}e^{-zt}\phi(t)=z\langle \bar{\bm{r}}_{\bm{0}}(z)\cdot\dot{\bm{r}}_{\bm{0}}(0)\rangle$
 of $\phi(t)$ satisfies
\begin{eqnarray}
\tilde{\phi}(z)&=&\frac{1}{(2\pi)^3}
\frac{
z\langle \hat{\bm{r}}_{\bm{k}}(0)\cdot\dot{\bm{r}}_{\bm{0}}(0)\rangle+\langle \dot{\hat{\bm{r}}}_{\bm{k}}(0)\cdot\dot{\bm{r}}_{\bm{0}}(0)\rangle
}{
z^2+4\omega_0^2\sum_{i=1}^3\sin^2\left(\frac{k_i}{2}\right)
}
\nonumber\\
&=&
\frac{3T z}{(2\pi)^3m}
\int d\bm{k} 
\frac{1}{z^2+4\omega_0^2\sum_{i=1}^3\sin^2\left(\frac{k_i}{2}\right) } ,
\label{phi(z)}\end{eqnarray}
where we have used $\langle \dot{\hat{\bm{r}}}_{\bm{k}}(0)\cdot\dot{\bm{r}}_{\bm{0}}(0)\rangle =3T/m$.
From the inverse Laplace transform of $\tilde{\phi}(z)$ we obtain the expression
\begin{equation}\label{phi(t)}
\phi(t)=\frac{3T}{m}\int \frac{d\bm{k}}{(2\pi)^3}\cos\left[2\omega_0t \sqrt{\sum_{i=1}^3\sin^2\left(\frac{k_i}{2}\right)} \right].
\end{equation}

\begin{figure}[th]
\begin{center}
 \begin{minipage}{0.47\textwidth}
  \includegraphics[width=0.8\textwidth]{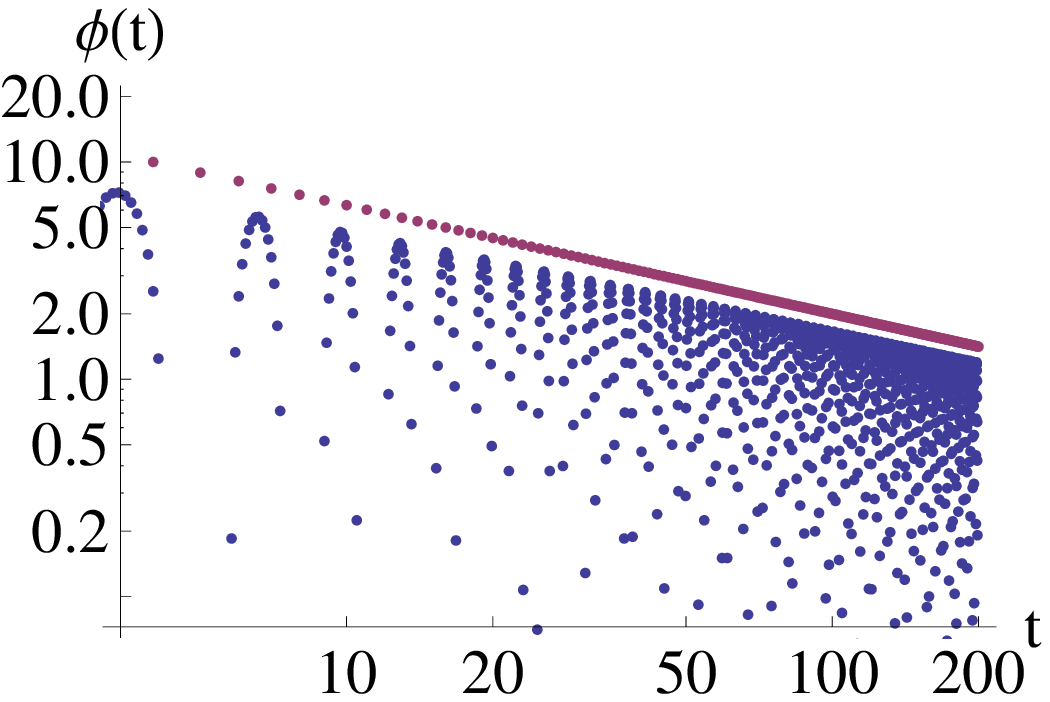}
  \caption{A log-log plot of the decay of VACF. Here the time is normalized by $\omega_0$ and the guide line represents
$20/\sqrt{t}$.}
\label{harm}
 \end{minipage}
\hspace*{3mm}
 \begin{minipage}{0.47\textwidth} 
  \includegraphics[width=0.8\textwidth]{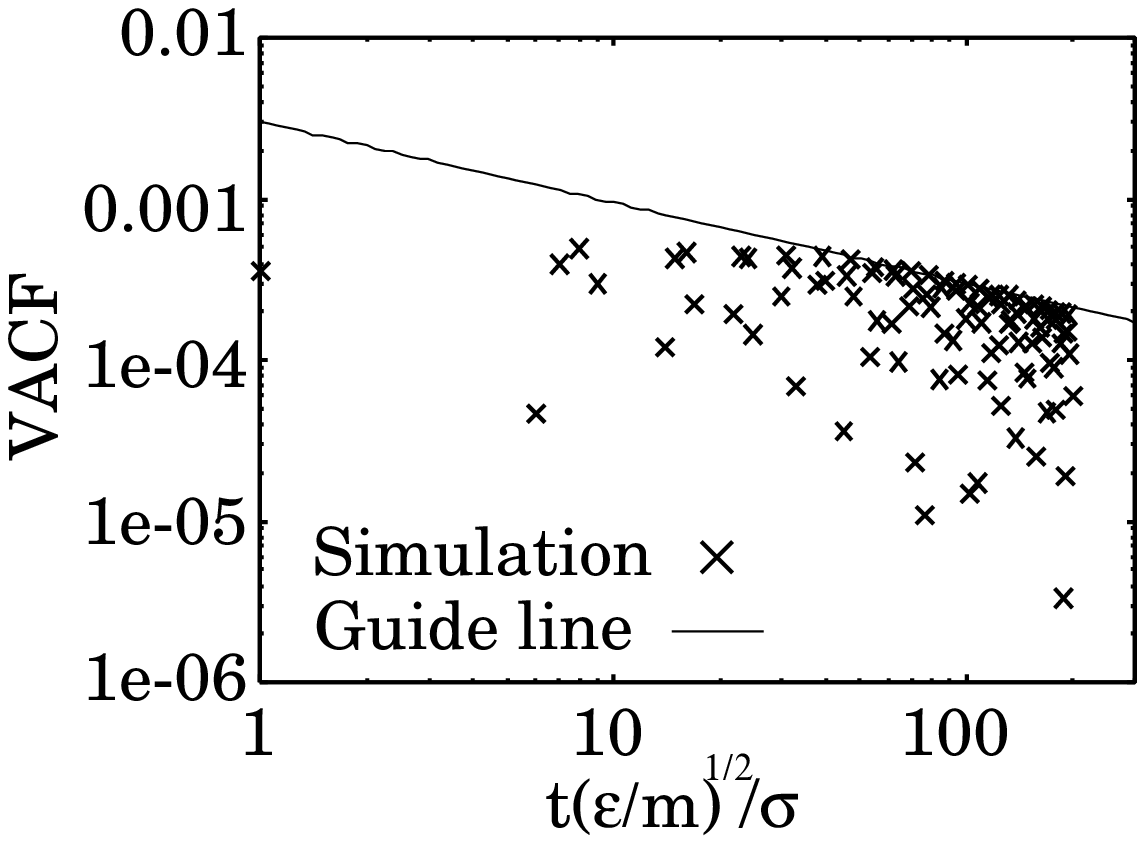}
  \caption{The comparison between VACF
   of the center of mass of the upper cluster (cross points) and fitting 
   line proportional to $1/\sqrt{t}$, where the time is dimensionless time for the MDS. }
  \label{vacf}
 \end{minipage}

\end{center}
\end{figure}

Since the direct integration  of (\ref{phi(t)}) is difficult and we are not interested in the detailed properties of the lattice, 
we may introduce the approximation $\sum_{i=1}^3\sin^2\frac{k_i}{2}\simeq \sum_{i=1}^3 k_i^2/4=k^2/4\simeq \sin^2\frac{k}{2}$, where 
$k\equiv\sqrt{\sum_{i=1}^3k_i^2}$.
Once we adopt such an approximation,
we obtain
\begin{equation}\label{phi}
\phi(t)\simeq \frac{3T}{2\pi^2m}\int_0^{\pi}dk k^2 \cos\left[2\omega_0t \sin\frac{k}{2}\right]
\end{equation} 
From the numerical integration of this expression, it is clearly to find $\phi(t)\sim 1/\sqrt{t}$ (see Fig. \ref{harm}).  
Indeed, when we put $\tau\equiv 2\omega_0t \sin k/2$, there is the relation $k^2dk=4dk-4d\tau/(\omega_0t)+O(k^4dk)$.
If we ignore the terms of $k^4dk$, we obtain the approximate relation
$I(t)\equiv \int_0^{\pi}dk k^2\cos[2\omega_0t \sin k/2]\simeq 4\pi J_0(2\omega_0 t)-4\sin (2\omega_0 t)/(\omega_0t)$.
From the asymptotic form of the Bessel function, we obtain $I(t)\simeq \sqrt{\pi/(\omega_0t)}\cos(2\omega_0t-\pi/4)$
for $\omega_0t\gg 1$.
This result is essentially the same as that for one-dimensional case $\phi(t)\propto J_0(2\omega_0t)$.
This time dependence of VACF can be observed in the direct simulation of nanoclusters 
in which each cluster is a 13 layer of the spherical cut of face-centered cubic (FCC) lattice
, {\it i.e.} 682 Lennard-Jones atoms (Fig.3). 
The setup of our simulation will be explained in the latter part. 

In spite of this extremely slow relaxation under the absence of $\hat{\gamma}(0)$,
we can approximately define the friction constant in the Langevin equation to describe the low frequency behaviors. 
Actually, if we adopt the approximation $\phi(t)\simeq \alpha\cos(\omega_0t)/\sqrt{\omega_0|t|}$ 
with a constant $\alpha$ for large $t$, 
substituting this into (\ref{fdr}) we obtain
\begin{equation}
(i\omega +\hat\gamma(\omega))^{-1}=
\alpha\frac{m}{T}\{\frac{1}{\sqrt{|2\omega_0-\omega|}}+\frac{1}{\sqrt{2\omega_0+\omega}}\}
\simeq \frac{\sqrt{2}\alpha m}{T\sqrt{\omega_0}}
(1+\frac{1}{4}\left(\frac{\omega}{2\omega_0}\right)^2+\cdots )
\end{equation}
for $\omega\ll \omega_0$. Thus, we may approximate 
$\gamma(t)=\int_{-\infty}^{\infty}d\omega/2\pi e^{i\omega t}\hat\gamma(\omega)$
by
\begin{equation}
\gamma(t)\simeq \Omega \delta(t)-\frac{d\delta(t)}{dt}
\end{equation}
where $\delta(t)$ is Dirac's delta function and $\Omega\equiv T\sqrt{\omega_0}/(\sqrt{2}\alpha m)$.
 Thus, the memory term can be approximated by
\begin{equation}
\int_{-\infty}^{t}dt'\gamma(t')\bm{v}(t')\simeq \Omega \bm{v}(t)+\frac{d\bm{v}}{dt}-\delta(0)\bm{v}(t),
\end{equation}
where the last term can be absorbed in the initial condition.
Finally, we obtain the effective Langevin equation for the low frequency behavior at $t\ne 0$ as
\begin{equation}
\frac{d\bm{v}}{dt}=-\frac{\Omega}{2}\bm{v}+\frac{\bm{\theta}}{2}+\frac{\bm{F}}{2M}
\end{equation}
which does not have any essential difference from the conventional Langevin equation. 
This may justify to use the quasi-static theory even when we 
consider a collision between nanoclusters of a uniform lattice system.
Indeed, once we accept to use the Langevin equation, it is straightforward to derive the quasi-static theory
of macroscopic collisions.\cite{kuwabara,brilliantov96,morgado}

It should be noted that the motion of the atom at the center of mass can be described by
the equation of motion for a harmonic oscillator. In order to use eq.(1), we need to introduce some tricks,
such as the mass difference, the contact with the other atoms and the nonlinearity. However, this argument
may be instructive to understand the basis of the Langevin equation from the mechanical point of view.

\section{Our numerical model}
Let us introduce our numerical model. Our model consists of 
two identical clusters. Each of them is the spherical cut from a 13 layered
face-centered cubic (FCC) lattice and consisted of $682$ ``atoms''.
When we simulate larger size of nanoclusters, the system is fluidized in the vicinity of surface,
 while the data for the smaller systems strongly depend on the specific orientation of impacts.
The details of system size dependence of the simulation will be reported elsewhere. 

The clusters have facets because of the small number of ``atoms'' 
(Fig. \ref{snap-LJ-rep}). 
All the ``atoms'' in each cluster are bounded  
by the Lennard-Jones potential $U(r_{ij})$ as 
\begin{equation}\label{LJ}
U(r_{ij})=4\epsilon\left\{\left(\frac{\sigma}{r_{ij}}\right)^{12} - a
               \left(\frac{\sigma}{r_{ij}}\right)^{6}\right\}, 
\end{equation}
where $r_{ij}$ is the distance between two ``atoms'', $i$ and $j$. 
The coupling coefficient of the attractive term $a$ is 
treated as a cohesive parameter  between atoms on the surfaces of one cluster and those on the surface of another, 
while the potential act on the atoms within the same cluster satisfies $a=1.0$. Here, 
we will consider collisions for the control parameters  $a=0$ and  $a=0.2$  between different clusters.\cite{awasthi}. 
In eq.(\ref{LJ}), $\epsilon$ is the energy constant and $\sigma$ is the lattice constant. 
When we regard the ``atom'' as argon, the values of the constants become 
$\epsilon=1.65\times10^{-21}\mathrm{J}$ and $\sigma=3.4$\AA, 
respectively.~\cite{rieth}
Henceforth, we label the upper and the lower clusters as 
cluster $C^{u}$ and cluster $C^{l}$, respectively. 
To reduce computational costs, we introduce the cut-off length $\sigma_{c}$ 
of the Lennard-Jones interaction as $\sigma_{c}=2.5 \sigma$. 

The procedure of our simulation is as follows. 
The initial velocities of the ``atoms'' in both $C^{u}$ and $C^{l}$ 
satisfy the Maxwell-Boltzmann distribution at the initial temperature $T$. 
The initial temperature is set to be $T=0.01 \epsilon$ or $T=0.02 \epsilon$ 
in most of our simulations.  
Sample average is taken over different sets of initial velocities 
governed by the Maxwell-Boltzmann velocity distribution for ``atoms''. 

To equilibrate the clusters, we adopt the velocity scaling 
method~\cite{haile,andersen} for 
$2000$ steps in the initial stage of simulations. 
We have checked the equilibration of the total energy 
in the initial relaxation process.
After the equilibration, 
we give translational velocities
to $C^{u}$ and $C^{l}$ at the relative separation  $\sigma_c$ between two clusters to make them collide against each other, 
where the initial colliding speed is achieved by the acceleration 
$g=0.01 \epsilon/(m \sigma)$ from a stationary state.
The relative speed of impact 
ranges from $V=0.02 \sqrt{\epsilon/m}$ to 
$V=0.07 \sqrt{\epsilon/m}$, 
which are less than the thermal velocity for one ``atom'' 
defined by $\sqrt{T/m}$, where $m$ is the mass of the ``atom''.

Numerical integration of the equation of motion for each atom 
is carried out by the second order symplectic integrator with 
the time step $dt=1.0 \times 10^{-2} \sigma/\sqrt{\epsilon/m}$. 
The rate of energy conservation, $|E(t)-E_{0}|/|E_{0}|$, 
is kept within $10^{-5}$, 
where $E_{0}$ is the initial energy of the system and $E(t)$ is the 
energy at time $t$.

We let the angle around $z-$axis, $\theta^{z}$, be $\theta^{z}=0$ 
when the two clusters are located in mirror-symmetric positions 
with respect to $z=0$. 
In most of our simulation, we adopt the data  at $\theta^{z}=0$. 
From our impact simulation for 
$\theta^{z}_{i}=\pi i/18 \hspace{1mm} (i=1,...,9)$ at 
$T=0.02\epsilon$ we have confirmed that the initial orientation 
does not crucially 
affect the restitution coefficient. 

\section{The results of our numerical simulation}

\subsection{The relaxation of velocity autocorrelation function}
At first, we have carried out the contact simulation for two identical nanoclusters 
contacting each other.
From our simulation, 
we verify that  
the Hertzian contact theory can be used without introduction of any fitting parameters\cite{kuninaka-preprint}. 
The details will be reported elsewhere. 
Another purpose of the contact simulation is to check whether 
eq.(\ref{phi})  can be used in our system.  
For this purpose, we make the two identical clusters contact each other under the mirror symmetric configuration,  
and equilibrate them at $T=0.03\epsilon$. 
After the equilibration, we leave 
those clusters, and record the time evolution of the velocity of 
$14$ atoms near the center of mass of the upper cluster, and collect
$50$ samples with different initial velocities for all the atoms, {\it i.e.} we average the data under 700 different samples to calculate VACF .


Figure \ref{vacf} is the result of VACF near the 
center of mass of the upper cluster in our simulation. 
The upper envelope line 
is given 
by $f(x)=0.0029 x^{-1/2}$, which
 is consistent  
with the theoretical prediction.

\subsection{The collision of two identical clusters}

In this subsection, we show the results of out simulations for colliding two 
identical nanoclusters. We mainly simulate the two cases for the interaction 
between different clusters: 
the completely repulsive case with $a=0.0$, and the weakly cohesive 
case with $a=0.2$, where $a$ is the cohesive parameter in eq. (\ref{LJ}) between different clusters. 

\begin{figure}[h]
\begin{center}
\includegraphics[width=.65\textwidth]{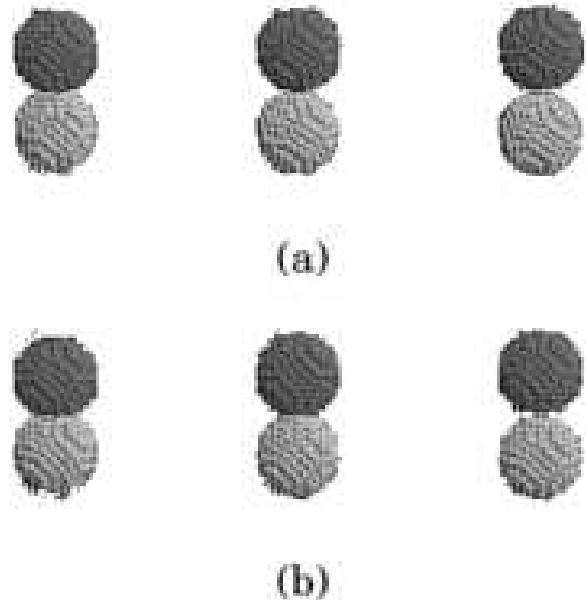}
\end{center}
\caption{Sequential snapshots of two colliding clusters in the cases of 
(a) a=0.0 and (b) a=1.0. }
\label{snap-LJ-rep}
\end{figure}

Let us show the sequential snapshots of two colliding 
clusters. Figure \ref{snap-LJ-rep} (a) and (b) show the collisional behavior 
in the case of $a=0.0$ and $a=1.0$, respectively.  
It should be noted that we demonstrate the case of $a=1.0$ to emphasize the difference between 
the noncohesive collision and the cohesive collision.
In Fig. \ref{snap-LJ-rep} (b),  we can observe 
the elongation of clusters along the $z-$axis before the separation, 
while we do not observe any elongation of 
clusters before the separation in Fig. \ref{snap-LJ-rep} (a). 
This elongation in Fig. \ref{snap-LJ-rep} (b) is the result of the cohesive interaction between two clusters.

\subsection{The relations between the restitution coefficient and the colliding speed}

\begin{figure}[th]
\begin{center}
 \begin{minipage}{0.42\textwidth} 
  \includegraphics[width=0.8\textwidth]{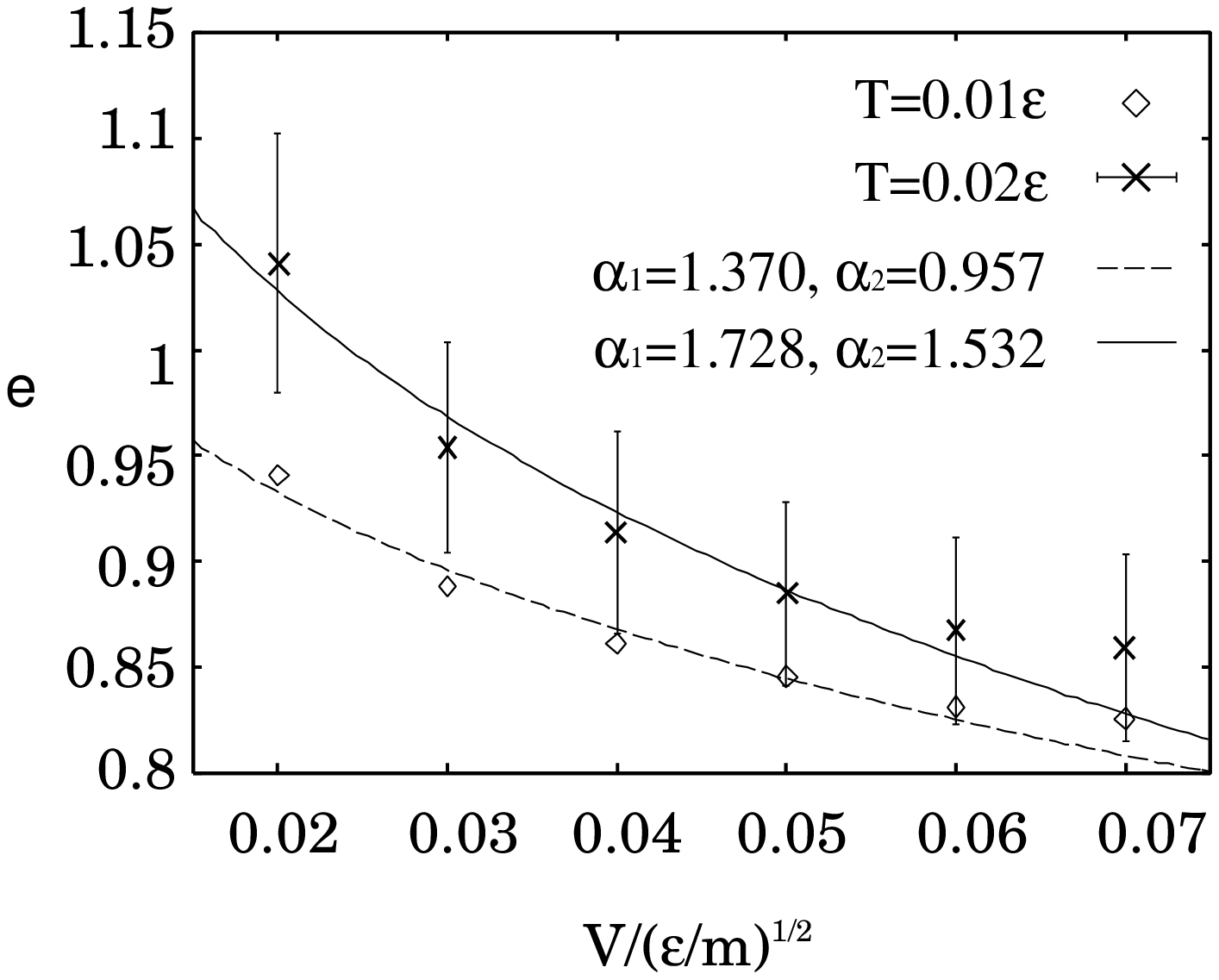}
  \caption{The relation between colliding speed and restitution 
   coefficient in the case of $a=0.0$ at
$T=0.01\epsilon$ and $T=0.02\epsilon$.}
  \label{fig2-2}
 \end{minipage}
\hspace*{3mm}
 \begin{minipage}{0.42\textwidth} 
\includegraphics[width=.8\textwidth]{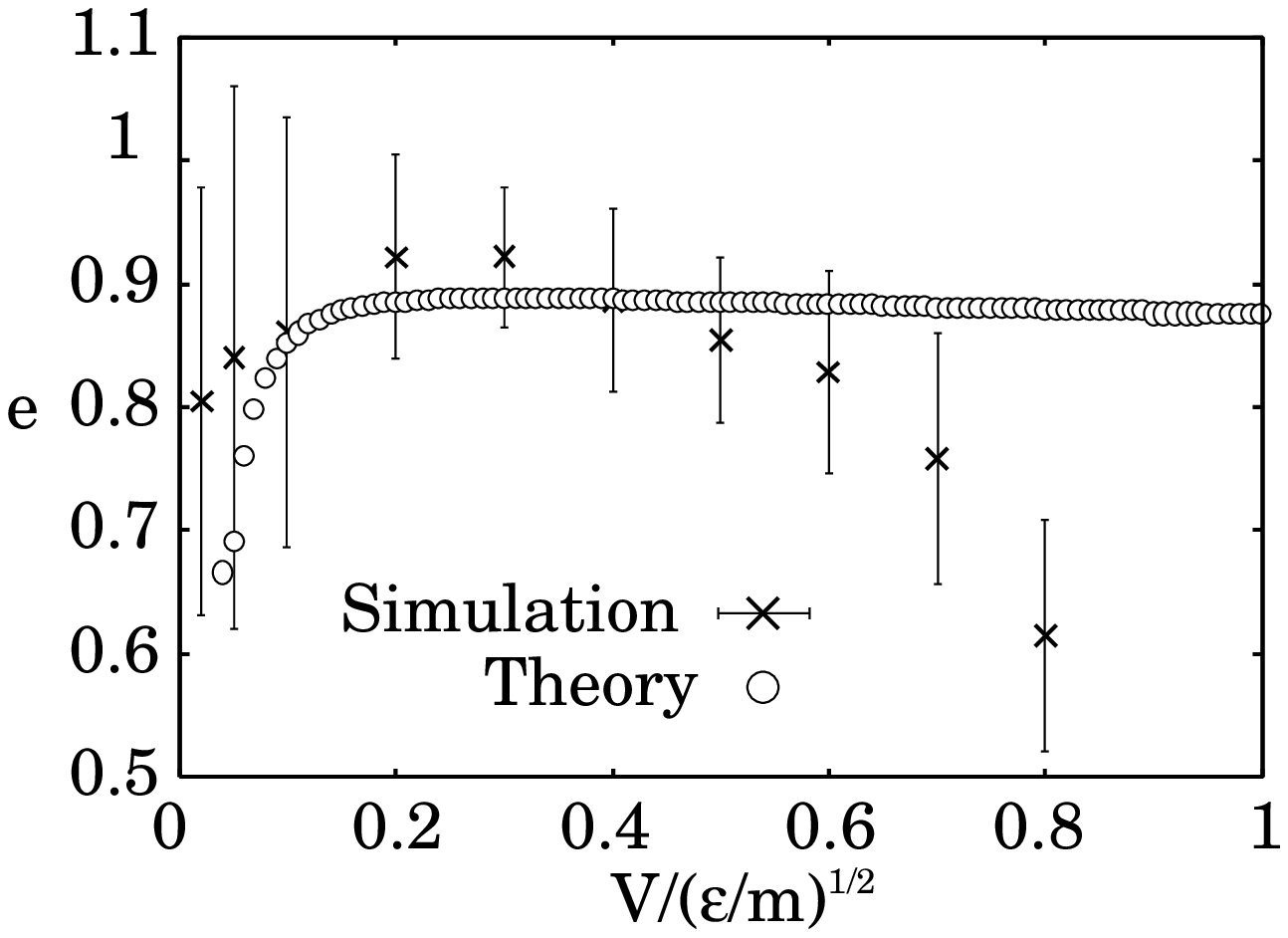}
\caption{The relation between colliding speed and restitution 
coefficient in the case of $a=0.2$. One cluster consists of 
682 atoms. The cross points and open circles are, respectively, the result of our simulation and  the theory in 
ref.\cite{brilliantov-adh}.}
\label{cor-adh}
 \end{minipage}
 \end{center}
\end{figure}

Here, we numerically investigate the relation between  
the restitution coefficient and the colliding speed. 
Figure \ref{fig2-2} shows the relation between the restitution coefficient $e$ and the relative speed of 
impact $V/ \sqrt{\epsilon/m}$ in purely repulsive collisions with $a=0$.
The initial configurations of two colliding clusters are assumed to be mirror symmetric.
The cross points and error bars in Fig.\ref{fig2-2} are, respectively,  the average and the standard deviation 
of $100$ samples for each colliding speed.  
From Fig. \ref{fig2-2}, we confirm that the restitution coefficient $e$ 
decreases with the increase of the colliding speed 
$V / \sqrt{\epsilon/m}$. When the colliding speed is 
$V = 0.02 \sqrt{\epsilon/m}$ at $T=0.02\epsilon$, 
the average of $e$ becomes $1.04$ 
which is slightly larger than unity.  
It is interesting that our result can be
fitted by the quasi-static theory 
of low-speed impacts $1-e\propto V^{1/5}
$\cite{kuwabara,brilliantov96,morgado} 
when the restitution coefficient
in the limit $V\to 0$ is replaced by a constant larger than unity. 
Indeed, the solid and the broken lines in Fig. \ref{fig2-2} are 
fitting curves of 
$e=\alpha_{1}-\alpha_{2} \left(V/ \sqrt{\epsilon/m}\right)^{1/5}$, 
where $\alpha_{1}$ and $\alpha_{2}$ depend on material constants of colliding 
bodies and $T$. 

We also briefly discuss the effect of the size dependence  
on the result. The results of our simulation for $N=433$,
 which is a 11 layered spherical cut of FCC lattice, cannot be
approximated by the quasi-static theory, where the restitution coefficient
seems to be almost independent of the colliding speed in the wide range of the impact speed. 
On the other hand,  
we cannot find any systematic relation between the restitution coefficient and the colliding speed in
the simulation for $N=1466$ which is a 17 layered spherical cut of FCC lattice.
This can be attributed 
to the melting on the surface of the cluster.
The details of the melting properties will be reported elsewhere.

Next, we investigate the weakly cohesive collisions with $a=0.2$ between those two clusters. 
Figure \ref{cor-adh} shows the relation between restitution coefficient 
and impact speed, where $100$ samples are taken 
for each colliding speed at the initial temperature $T=0.02\epsilon$. 
When there is the cohesive interaction between two colliding clusters, the relation 
has a peak as suggested by Brilliantov {\it et al.}\cite{brilliantov-adh}. 
In the figure, the open circles are the numerical results obtained by solving the equation 
developed by Brilliantov {\it et al.}\cite{brilliantov-adh}.
To solve this equation, we evaluate the values
$\gamma\simeq 0.026\sqrt{\epsilon m/\sigma}$ 
from the calculation of the attractive interaction between 
two clusters.\footnote{The surface tension $\gamma$ can be calculated from the attractive potential. 
The method of our evaluation will be reported elsewhere.}
The theoretical result in  Fig. \ref{cor-adh} suggests 
that the restitution coefficient is insensitive to the colliding speed for the large colliding speed,
though the restitution coefficient slightly decreases with the increment of the colliding speed. 

\subsection{The frequency distribution functions of the restitution coefficient}
\begin{figure}[th]
\begin{center}
 \begin{minipage}{0.47\textwidth}
  \includegraphics[width=0.8\textwidth]{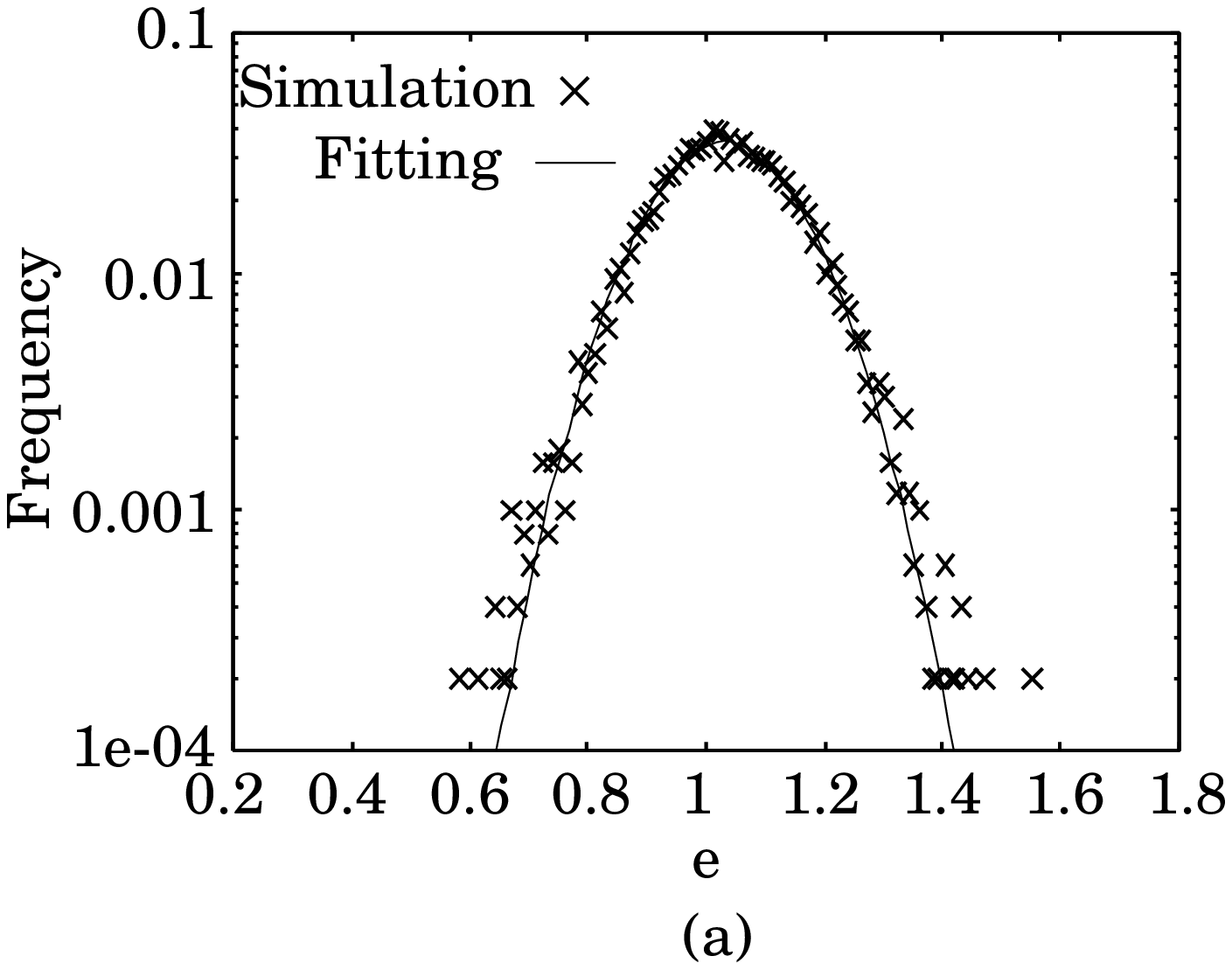}
 \end{minipage}
\hspace*{3mm}
 \begin{minipage}{0.47\textwidth} 
  \includegraphics[width=0.8\textwidth]{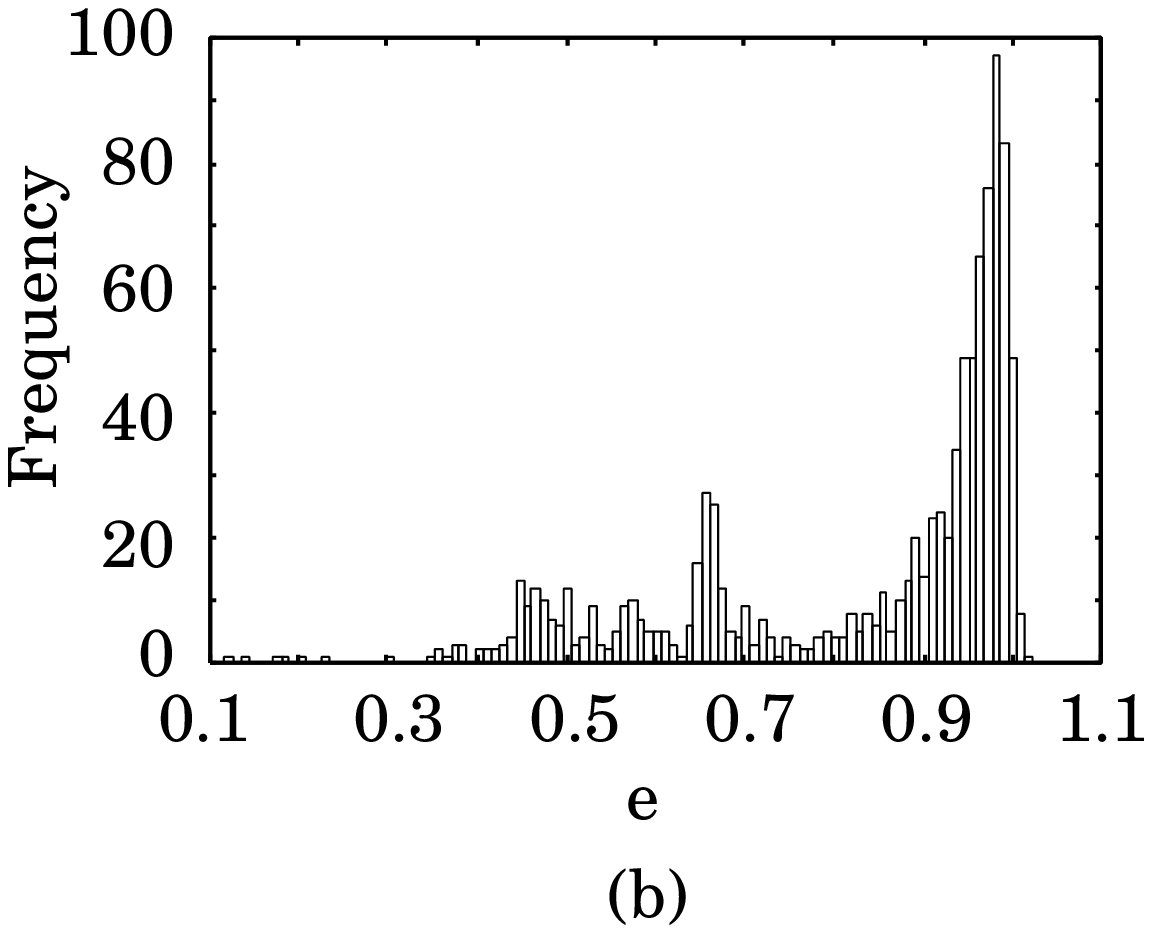}
 \end{minipage}
\caption{Histograms of the restitution coefficients for (a) $a=0.0$,
 $V=0.02\sqrt{\epsilon/m}$, 
and (b) $a=0.2$, $V=0.1\sqrt{\epsilon/m}$. 
The guide line in the left figure is the Gaussian fitting of the data.}
\label{hist-a0_2-v0_1}
\end{center}
\end{figure}

Here, we show our numerical results on the frequency distribution function of the restitution coefficient, which strongly depends on the cohesive parameter. 
Figure \ref{hist-a0_2-v0_1} shows histograms of the restitution coefficients 
for both the purely repulsive collisions and the cohesive collisions $a=0.2$.
When there is no cohesive interaction between the two clusters, the frequency distribution function is roughly represented by the Gaussian distribution
function. On the other hand, the frequency distribution function is irregular when the cohesion exists. It is notable that
the anomalous events for $e$ to exceed the unity becomes rare when there is the attractive interaction between clusters, though a few percent of the collisions
still exhibit the anomalous impacts. This is because two clusters are
coalesced with each other in the slow impacts. Therefore, the frequency
distribution function for $a=0.2$ has a steep peak near $e=1$. 

In Fig.\ref{hist-a0_2-v0_1}(b), there are the first and the second peaks 
around $e=0.4$ and $e=0.65$, respectively. 
The collisional modes observed around these peaks are the rotational bounces 
after the collisions, while the most of bounces are not associated with rotations 
in the vicinity of  the third peak around $e=1$. 
It is reasonable that the excitation of macroscopic rotation lowers 
the translational energy to decrease the restitution coefficient. 

\begin{figure}[h]
\begin{center}
\begin{minipage}{0.47\textwidth} 
\includegraphics[width=.8\textwidth]{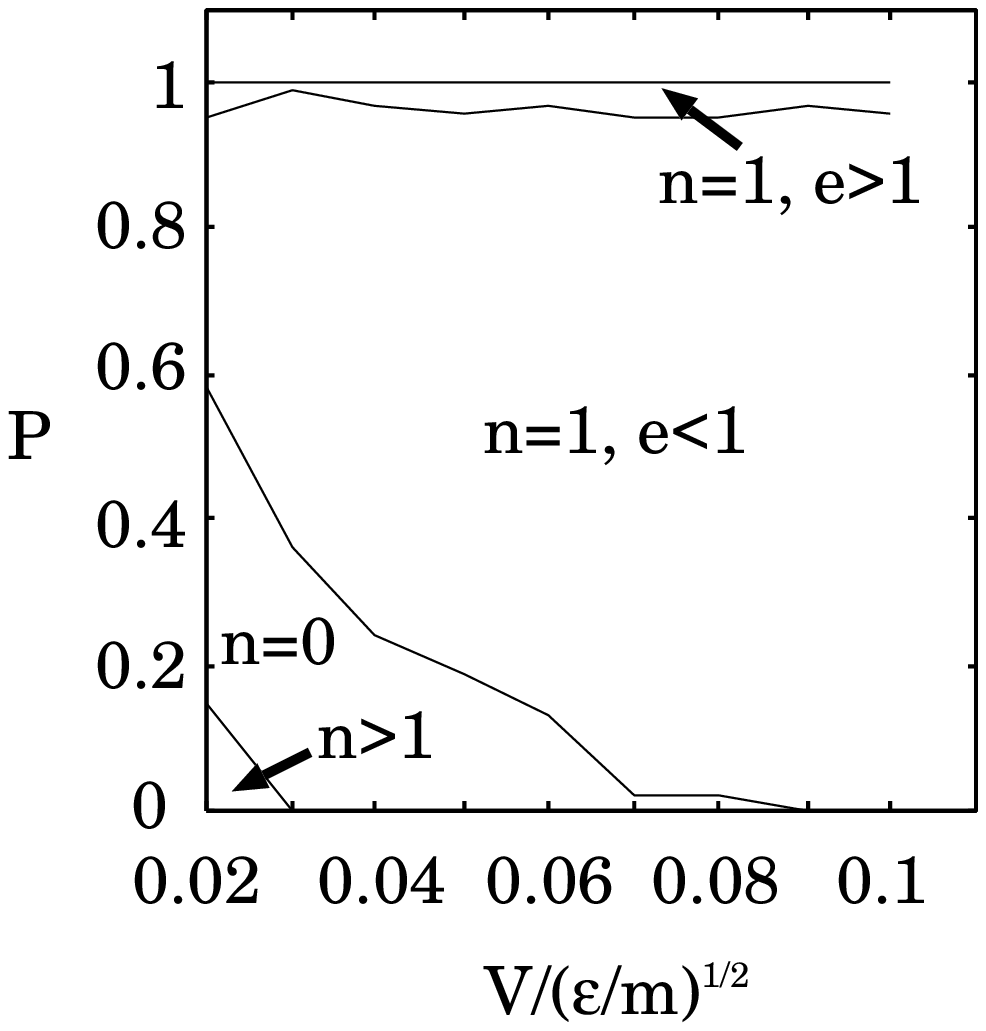}
\caption{Probabilities classified by collision modes observed 
in cohesive collision with $a=0.2$. }
\label{prob2}
\end{minipage}
\hspace*{3mm}
\begin{minipage}{0.47\textwidth} 
\includegraphics[width=.8\textwidth]{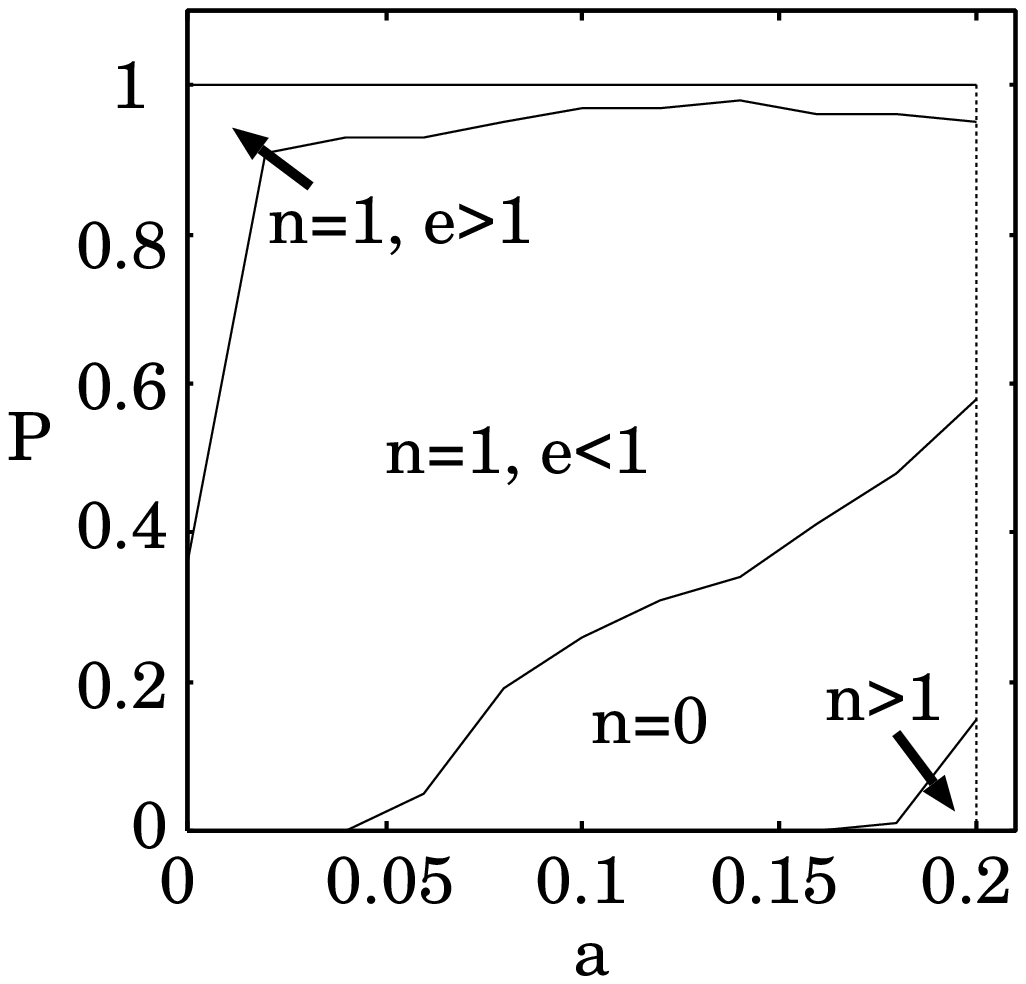}
\caption{The probability classified by four modes at $V=0.02\sqrt{\epsilon/m}$.}
\end{minipage}
\end{center}
\end{figure}

For  cohesive collisions, we can categorize the rebound behaviors of the colliding clusters into four patterns (see 
Fig. \ref{prob2}):
   (a) $n=0$ (complete adhesion), (b) $n>1$, 
(c) $n=1$ and $e<1$, and (d) $n=1$ and $e>1$, where $n$ is the number of collisions in each impact process.
The collision with $n>1$ can take place, when the attractive interaction between the colliding clusters exists. Indeed,
if the rebound speed is not large enough, the rebounded clusters are attracted to have the second collision. 
We call the case with $e>1$ and $n=1$ the anomalous impact, but there are some other characteristic collisions as can be seen in Fig. \ref{prob2}. 

Similarly, we categorize the collisions into four groups as a function of the cohesive parameter under the fixing
colliding speed $V=0.02\sqrt{\epsilon/m}$ (Fig. 8).
It is obvious that there are two categories, (c) and (d),
 in noncohesive collisions, while the probability to occur (a) or (b) increases as $a$ increases.
It is interesting that Fig.8 is almost the mirror symmetric one of Fig. 7. This fact suggests that the cohesive parameter plays a role of the impact speed.
The relation between the impact speed and the cohesive parameter will be discussed elsewhere.

\section{Discussion and conclusion}

In this paper, we study collisions of nanoclusters which are thermally activated. We also discuss the effects of cohesive force between 
the colliding two clusters. Although the results are preliminary, we believe that our paper includes some potentially important results 
for the nanoscience.  Let us briefly discuss our results. An anomalous impact with  $e>1$ occurs with a finite probability even for realistic situations (see Fig. 8).
This is an important indication, though  the cohesive force between the colliding clusters suppresses such events
in the low speed collisions. 
We also find an interesting similarity in the roles of the impact speed and the cohesive parameter (Fig. 8).
It is  more interesting that the cluster is fluidized when the cluster is large. There is capillary instability at the surface of the large cluster,
because the influence of the attractive binding force from the center of mass is weaker, the size of the cluster is larger.  The quantitative
discussion will be discussed elsewhere.

In conclusion, we study the impact of thermally activated nanoclusters numerically. We confirm that VACF satisfies
$\phi(t)\sim 1/\sqrt{t}$. The restitution coefficient seems to be consistent with the quasi-static theory when
there is no attractive interaction between the two colliding clusters, while the restitution coefficient has a peak 
at a finite value of the impact speed, when the attractive interaction exists.
The anomalous impacts which have $e>1$ commonly take place  in purely repulsive collisions, while 
such an impacts become rare in cohesive collisions. The frequency distribution function satisfies Gaussian for purely repulsive collisions and
has some peaks in cohesive collisions.

\section*{Acknowledgements}
{\it The authors are deeply grateful to N. V. Brilliantov to give them the opportunity to present this work.
This work is partially supported by
Ministry of Education, Culture, Sports, Sciences and Technology (MEXT) Japan (Grant No. 18540371).
}

\bigskip

\address{Hisao Hayakawa, Yukawa Institute for Theoretical Physics, Kyoto University, Kitashirakawa-oiwakecho, 
Sakyo-ku, Kyoto 606-8502, Japan}
\address{Hiroto Kuninaka, Department of Physics, Chuo University, Bunkyo-ku, Tokyo 112-8551, Japan}

\end{document}